\documentclass[aps,twocolumn,superscriptaddress,amsmath,amssymb,nofootinbib]{revtex4-1}
\usepackage{graphicx}  
\usepackage{dcolumn}   
\usepackage{bm}        
\usepackage{verbatim}   
\usepackage{gensymb}
\usepackage{epstopdf}
\usepackage{footmisc}
\usepackage{natbib}
\usepackage{color}
\usepackage{hyperref}

\hypersetup{colorlinks, linkcolor={blue}, citecolor={blue}, urlcolor={blue}}
\graphicspath{{./}{./figs/}}

\newcommand{\lllangle}{\langle\!\langle\!\langle}
\newcommand{\rrrangle}{\rangle\!\rangle\!\rangle}

\newcommand{\rucl}{$\alpha$-RuCl$_3$}

\begin{document}

\title{Field-induced transitions of the Kitaev material $\alpha$-RuCl$_3$ probed by thermal expansion and magnetostriction}

\author{S. Gass}
\affiliation{Institut f\"ur Festk\"orperforschung, Leibniz IFW Dresden, 01069 Dresden, Germany}
\affiliation{Institut f\"ur Festk\"orper- und Materialphysik and W\"urzburg-Dresden Cluster of Excellence
ct.qmat, Technische Universit\"at Dresden, 01062 Dresden, Germany}
\author{P. M. C\^onsoli}
\affiliation{Institut f\"ur Theoretische Physik and W\"urzburg-Dresden Cluster of Excellence
ct.qmat, Technische Universit\"at Dresden, 01062 Dresden, Germany}
\affiliation{Instituto de F\'isica de S\~ao Carlos, Universidade de S\~ao Paulo, C.P. 369,
S\~ao Carlos, SP, 13560-970, Brazil}
\author{V. Kocsis}
\author{L. T. Corredor}
\affiliation{Institut f\"ur Festk\"orperforschung, Leibniz IFW Dresden, 01069 Dresden, Germany}
\author{P. Lampen-Kelley}
\author{D. G. Mandrus}
\affiliation{Materials Science and Technology Division, Oak Ridge
National Laboratory, Oak Ridge, Tennessee 37831, USA}
\affiliation{Department of Materials Science and Engineering,
University of Tennessee, Knoxville, Tennessee 37996, USA}
\author{S. E. Nagler}
\affiliation{Neutron Scattering Division, Oak Ridge National
Laboratory, Oak Ridge, Tennessee 37831, USA}
\author{L. Janssen}
\author{M. Vojta}
\affiliation{Institut f\"ur Theoretische Physik and W\"urzburg-Dresden Cluster of Excellence
ct.qmat, Technische Universit\"at Dresden, 01062 Dresden, Germany}
\author{B. B\"{u}chner}
\affiliation{Institut f\"ur Festk\"orperforschung, Leibniz IFW Dresden, 01069 Dresden, Germany}
\affiliation{Institut f\"ur Festk\"orper- und Materialphysik and W\"urzburg-Dresden Cluster of Excellence
ct.qmat, Technische Universit\"at Dresden, 01062 Dresden, Germany}
\author{A. U. B. Wolter}
\email{a.wolter@ifw-dresden.de} \affiliation{Institut f\"ur Festk\"orperforschung, Leibniz IFW Dresden, 01069 Dresden, Germany}

\date{\today}


\begin{abstract}
High-resolution thermal expansion and magnetostriction measurements were performed on single crystals of {\rucl} in magnetic fields applied parallel to the Ru-Ru bonds. The length changes were measured in the direction perpendicular to the honeycomb planes. Our data show clear thermodynamic characteristics for the field-induced phase transition at the critical field $\mu_\mathrm{0}H_{\mathrm c1} = 7.8(2)$\,T where the antiferromagnetic zigzag order is suppressed. At higher fields, a kink in the magnetostriction coefficient signals an additional transition or crossover around $\mu_\mathrm{0}H_{\mathrm c2} \approx 11$\,T. The extracted Gr\"uneisen ratio shows typical hallmarks for quantum criticality near $H_{\mathrm c1}$, but also displays anomalous behavior above $H_{\mathrm c1}$. We compare our experimental data with spin-wave calculations employing a minimal Kitaev-Heisenberg model in the semiclassical limit. Most of the salient features are in agreement with each other, however, the peculiar features in the region above $H_{\mathrm c1}$ cannot be accounted for in our semiclassical modelling and hence suggest a genuine quantum nature. We construct a phase diagram for {\rucl} in a magnetic field along the Ru-Ru bonds, displaying a zigzag ordered state below $H_{\mathrm c1}$, a quantum paramagnetic regime between $H_{\mathrm c1}$ and $H_{\mathrm c2}$, and a semiclassical partially polarized state above $H_{\mathrm c2}$.
\end{abstract}

\maketitle

\section{Introduction}

The search for realizations of topological quantum spin liquids (QSLs) has generated a tremendous excitement, for both fundamental reasons and potential applications, e.g., in quantum information processing~\cite{Nayak2008}. QSLs are characterized by long-range entanglement, topological order and associated ground-state degeneracies, as well as fractionalized quasiparticles. Kitaev's spin-1/2 model on the honeycomb lattice \cite{Kitaev2006} is a paradigmatic example for a QSL because it uniquely combines exact solvability in terms of Majorana fermions and experimental relevance~\cite{Jackeli2009,Chaloupka2010,Trebst2017,Banerjee2017}.

One of the prime candidates to realize Kitaev magnetism is the compound \rucl: It is a $J_{\rm eff}$ = 1/2 Mott insulator with a layered structure of edge-sharing RuCl$_6$ octahedra arranged in a honeycomb lattice \cite{Plumb2014,Sears2015,Johnson2015,Majumder2015,Kubota2015,Sinn2016,Ziatdinov2016,Weber2016}. While {\rucl} displays magnetic long-range order of so-called zigzag type, a moderate in-plane magnetic field suppresses the magnetic order, resulting in a paramagnetic state whose nature has been debated~\cite{Yadav2016,Leahy2017,Kasahara2018}.
By now, the existence of a quantum spin-liquid regime in {\rucl} in a window of applied magnetic field is suggested by a number of experimental results, such as an excitation continuum in neutron scattering~\cite{Do2017,Banerjee2018,Balz2019}, in Raman scattering~\cite{Sandilands2015}, as well as in microwave/terahertz absorption measurements~\cite{Wang2017,Wellm2018}, and, most prominently, an approximately half-quantized thermal Hall conductivity~\cite{Kasahara2018,Yokoi2020}. The latter has been associated with the presence of a chiral Majorana edge mode, characteristic of a Kitaev spin liquid in applied magnetic field \cite{Cookmeyer2018,Vinkleraviv2018}. Theoretically, a field-induced spin liquid has been discussed for microscopic models relevant to {\rucl}~\cite{Jiang2019,Gordon2019,Kaib2019}.

However, the structure of the field-temperature phase diagram of {\rucl} is not settled: The experiments of Refs.~\onlinecite{Kasahara2018,Balz2019,Yokoi2020} suggest the existence of at least three low-temperature phases, i.e., a spin-liquid phase sandwiched between the zigzag and high-field phases. Yet clear-cut thermodynamic evidence for a transition between the spin-liquid and high-field phase is lacking, perhaps with the exception of a signature in the magnetocaloric effect \cite{Balz2019}. Moreover, the spin-liquid signatures have not been traced to very low temperatures, hence they may as well represent a quantum critical regime instead of a stable phase.

In this paper, we report a thorough dilatometric study of {\rucl} in in-plane magnetic fields up to $14$\,T and temperatures down to $2.4$\,K. Thermal expansion (TE) and magnetostriction (MS) represent thermodynamic properties governed by magnetoelastic coupling, enabling us to study the nature of the different phase transitions and possible critical behavior. We confirm the field-induced suppression of long-range order at a critical field of $\mu_\mathrm{0}H_{\mathrm c1} = 7.8(2)$\,T and provide strong thermodynamic evidence for quantum critical behavior at $H_{\mathrm c1}$ by analyzing the Gr\"uneisen ratio. This is confirmed by our MS data, which moreover displays signatures of an additional weak first-order transition or crossover around $\mu_\mathrm{0}H_{\mathrm c2} \approx 11$\,T.
A comparison of our experimental data to semiclassical calculations in a minimal spin model yields qualitative agreement for many features, but also hints at additional physics between $H_{\mathrm c1}$ and $H_{\mathrm c2}$ beyond semiclassics. On basis of our experimental data we conjecture a field-temperature phase diagram of {\rucl} for in-plane fields along the Ru-Ru bonds containing three distinct low-temperature regimes.

\section{Experiments}
\subsection{Methods}

High-quality single crystals of {\rucl} with a thickness of $\sim 1$\,mm were grown using a vapor-transport technique~\cite{Banerjee2017}.  Via angular-dependent magnetization measurments~\cite{janssen2017,Lampen-Kelley2018} the sample was properly aligned to ensure that the magnetic field was applied parallel to the Ru-Ru bond direction.
This is the field direction where the additional ordered phase found in Ref.~\onlinecite{kelley2018b} is absent or very narrow.

The linear TE and MS of {\rucl} were determined by using a custom-built capacitive dilatometer with a parallel-plate system consisting of two separately aligned capacitor plates, which detect changes of the uniaxial sample length $\Delta L_i$. The sample is clamped between one of the plates and the frame of the dilatometer, and thus is exposed to a small force via the springs of the dilatometer. Note that this force in combination with the van-der-Waals bonds between the honeycomb planes of {\rucl} leads to an irreversible mechanical deformation of the sample along $ab$. Thus, all TE and MS studies were performed for the configuration $\Delta \vec L \parallel c^*$, $\Delta \vec L \perp \vec H$ for this van-der-Waals bonded material, with $c^*$ being perpendicular to the crystallographic $ab$ plane. For the TE the temperature $T$ was swept from 3\,K to 300\,K using sweep rates between 0.03\,K/min and 0.2\,K/min. The MS was measured at constant temperatures between 2.4\,K and 10\,K and slowly sweeping the magnetic fields from 0\,T to 14\,T (sweep rates of 0.01\,T/min and 0.03\,T/min). A correction for the TE of the dilatometer itself has been applied using high-purity Cu reference samples. Measurements were performed on two different single crystals with a thickness of $\sim 1.0$\,mm (sample~\#1) and $\sim 0.8$\,mm (sample~\#2).

Specific-heat measurements under applied magnetic fields $\vec H \parallel$ Ru-Ru bonds up to $14$\,T were performed on the same single crystal used for the dilatometry measurements (sample \#1). For the experiments a heat-pulse relaxation method was used in a Physical Property Measurement System (PPMS, Quantum Design). In order to obtain the intrinsic specific heat of \rucl, the temperature- and field-dependent addenda were subtracted from the measured specific-heat values in the sample measurements.
In order to estimate the phononic contribution, the specific heat of the non-magnetic structural analog compound RhCl$_3$ in polycrystalline form was measured~\cite{Wolter2017}.

\subsection{Thermal expansion}

\begin{figure}[tb]
\includegraphics[scale=0.95]{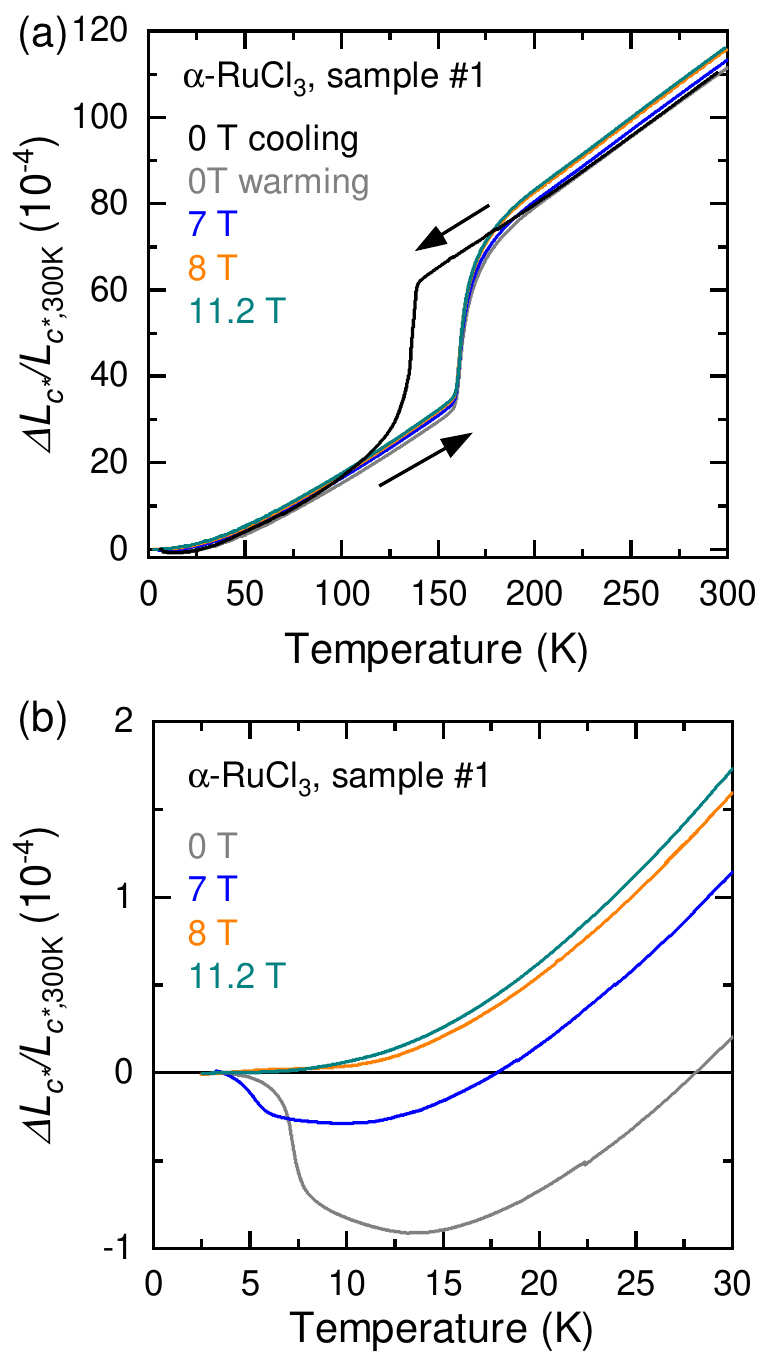}
\caption{\label{DeltaL}
Normalized linear TE of {\rucl} perpendicular to the $ab$ plane, $\Delta L_{c^*}/L_{c^*,300\,\text{K}}$, as function of temperature for both zero and finite applied magnetic fields $\vec H \parallel$ Ru-Ru bonds;
(a) full temperature interval up to $300$\,K,
(b) low-temperature region up to $30$\,K.
}
\end{figure}

In Fig.~\ref{DeltaL} the normalized linear TE measured along the $c^*$ direction,
\begin{equation}
\label{eq:thermalexpansion} \frac{\Delta L_{c^*}(T, \mu_0 H)}{L_{c^*} (300\,\text{K},0\,\text{T})} \equiv \frac{L_{c^*}(T,\mu_0 H)-L_{c^*}(T_{\rm ref},0\,{\rm T})}{L_{c^*}(300\,\text{K},0\,\text{T})},
\end{equation}
of {\rucl} is depicted for zero field as well as for some representative in-plane magnetic fields $\vec H \parallel$ Ru-Ru bonds. $\Delta L_{c^*}$ is the measured length change along the $c^*$ direction, which is then normalized to the sample length at room temperature. $T_{\rm ref} = 3.6$\,K represents a minimum reference temperature at which $L_{c^*}$ was measured for every data set and thus our data refer to, i.e., the TE is zero at $T_{\rm ref}$ and vanishing field.

The zero-field $c^*$-axis TE is rather large, as expected for weakly bonded van-der-Waals materials, Fig.~\ref{DeltaL}(a). Further, the overall TE is decreasing upon lowering temperature, which is in line with an overall shrinking of the lattice constants compared to room temperature~\cite{Johnson2015,Park2016}. Interestingly, a step-like feature is seen in our TE data at around $T_\mathrm{s,c} \approx 137$\,K upon cooling, clearly indicating a first-order structural transition. The transition is strongly hysteretic, with $T_\mathrm{s,w} \approx 161$\,K upon warming. It likely corresponds to a change from a high-temperature monoclinic $C2/m$ structure to a low-temperature rhombohedral $R\bar 3$ structure \cite{Park2016}. Note that $T_\mathrm{s,c}$ and thus also the hysteresis upon this transition strongly depend on the used temperature sweep rate.

Looking at the details at low temperatures in zero field, Fig.~\ref{DeltaL}(b), the overall shrinking of the $c^*$-axis TE is followed by a broad minimum at $\sim 14$\,K and a subsequent expansion of the $c^*$ axis down to lowest temperatures. Furthermore, a sharp kink is clearly discernable at the antiferromagnetic transition temperature $T_\mathrm{N} = 7.2(1)$\,K. While the TE for $T \gtrsim$ 100~K is not particularly sensitive to the application of an in-plane magnetic field, the low-temperature TE changes dramatically up to the critical field $\mu_\mathrm{0}H_{\mathrm c1} = 7.8(2)$\,T at which the kink signalling the antiferromagnetic transition is finally completely suppressed. For fields larger than $H_{\mathrm c1}$ a positive TE is discernable.

Overall, our TE data are in good agreement with x-ray diffraction and former zero-field TE experiments~\cite{Park2016,He2018}, showing a rearrangement of the unit cell of {\rucl} both at the structural and the antiferromagnetic phase transitions, and thus also a coupling of the lattice and spin degrees of freedom in our compound.

\begin{figure}[tb]
\includegraphics[scale=0.95]{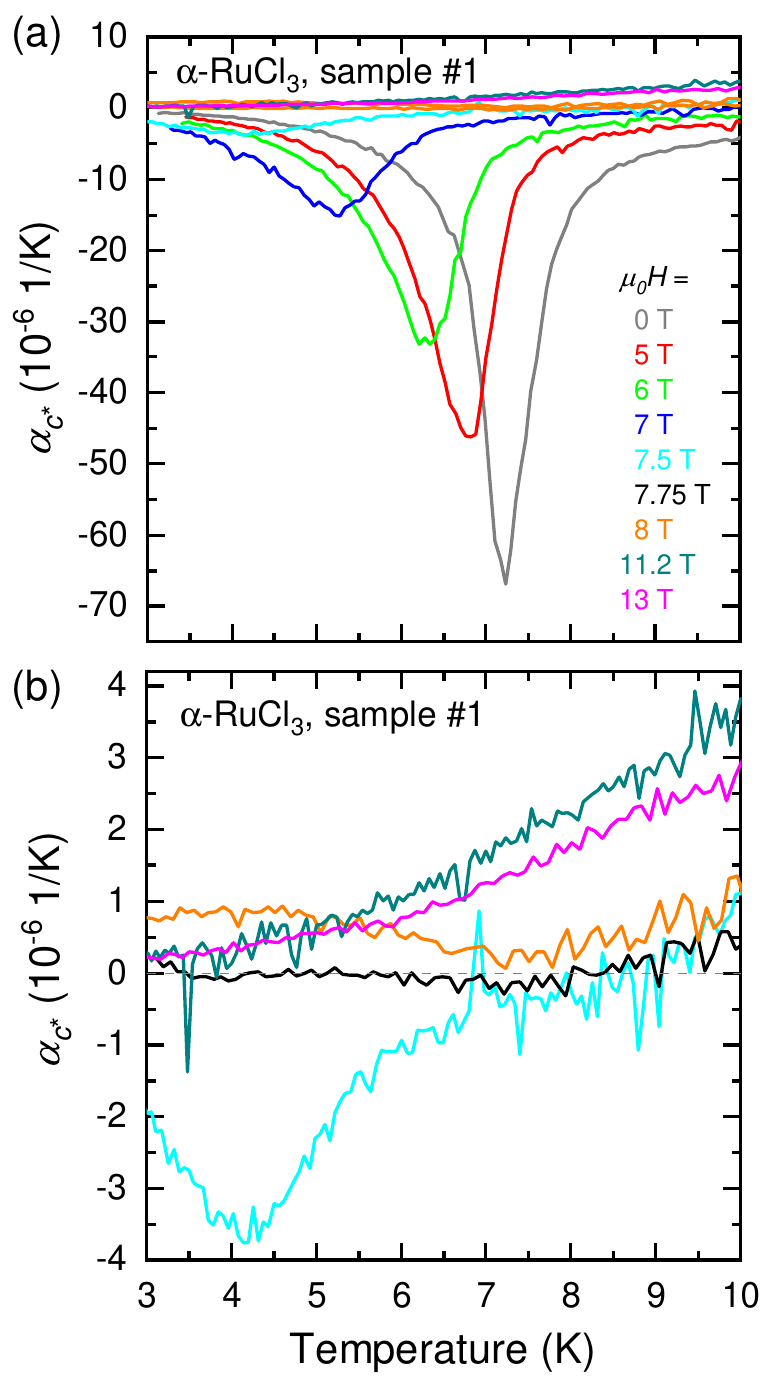}
\caption{\label{alpha}
The linear TE coefficient perpendicular to the $ab$ plane, $\alpha_{c^*}$, of {\rucl} for zero and applied magnetic fields $\vec H \parallel$ Ru-Ru bonds. (a) General overview of $\alpha_{c^*}$ with the focus on the ordered antiferromagnetic zigzag transition and its shift as function of the applied field. (b) A zoom of the data for
applied fields close to the critical field $\mu_\mathrm{0}H_{\mathrm c1}$ = 7.8(2)~T.}
\end{figure}

In order to better resolve the magnetic phase transitions, we also analyze the linear TE coefficient along the $c^*$ direction,
\begin{equation}
\label{eq:alpha} \alpha_{c^*} =
\frac{\partial}{\partial T}\frac{\Delta L_{c^*}(T,\mu_0 H)}{L_{c^*} (300\,\text{K},0\,\text{T})}
\end{equation}
Anomalies in $\alpha_{c^*}(T)$ typically correspond to phase transitions.
Low-temperature results for {\rucl} are shown in Fig.~\ref{alpha}(a). At zero field the sharp peak signifies a single phase transition at $T_\mathrm{N} = 7.2(1)$\,K. With increasing field the peak broadens, reduces in magnitude, and shifts to lower temperatures, until it disappears at $\mu_0H_{\mathrm c1} = 7.8(2)$\,T. Given the agreement with other probes, we conclude that this peak represents the magnetic transition into the zigzag phase. It highlights that the low-$T$ contributions to $\alpha_{c^*}$ are primarily magnetic, and also underlines the quality of our single crystals, with a dominant ABC stacking of the hexagonal layers along $c^*$.

Fig.~\ref{alpha}(b) displays a magnified region of Fig.~\ref{alpha}(a) for fields near and above $H_{\mathrm c1}$. The low-$T$ linear TE shows a sign change close to the critical field~$H_{\mathrm c1}$; at $H_{\mathrm c1}$ the TE coefficient is tiny up to about $8$\,K, indicating that the phonon contribution is small in this temperature regime. For fields of $11.2$\,T and $13$\,T, $\alpha_{c^*}(T)$ is positive and monotonic. In this high-field regime, the magnitude of $\alpha_{c^*}$ decreases with increasing field, consistent with an increasing magnetic excitation gap in the polarized high-field phase as observed by various methods, such as nuclear magnetic resonance and thermal conductivity~\cite{Baek2017,Hentrich2018}.
Interestingly, the data at $8$\,T are anomalous in that $\alpha_{c^*}$ shows a non-monotonic $T$ dependence, suggesting the existence of a distinct intermediate-field region between the zigzag and high-field phases, as recently also observed with other techniques~\cite{Kasahara2018,Yokoi2020,Balz2019}.

\subsection{Gr\"uneisen ratio}
\label{subsec:exp-grueneisen}

The linear TE coefficient is proportional to the derivative of the entropy with respect to uniaxial pressure along the $c^*$ axis, $\partial S/\partial p_{c^*}$, as discussed in detail below (see Sec.~\ref{sec:th-methods}). Therefore, vanishing $\alpha_{c^*}$ near $H_{\mathrm c1}$ indicates a maximum of the magnetic contribution to the entropy, $S_{\rm mag}$, at the critical field. In fact, such an entropy accumulation is predicted to occur near a continuous quantum phase transition~\cite{Zhu2003,Garst2005}, and can be identified experimentally by measuring the Gr\"uneisen ratio, commonly defined as the ratio between the magnetic contributions to the volume TE coefficient and the specific heat $C_{p,\mathrm{mag}}$,
\begin{equation}
\label{eq:grueneisen}
\Gamma = V_{\mathrm m}\frac{\alpha_{\rm mag}}{C_{p,\rm mag}} = -\frac{(\partial S_{\rm mag}/\partial p)_T}{T(\partial S_{\rm mag}/\partial T)_p},
\end{equation}
where $V_{\mathrm m} \simeq 55.9\,\text{cm}^3/\text{mol}$ is the molar volume~\cite{Johnson2015,Park2016,Cao2016}. $\Gamma$ displays characteristic divergencies \cite{Zhu2003} upon approaching a pressure-driven quantum critical point (QCP), and both $\Gamma$ and $\alpha_{\rm mag}$ change sign near a QCP as a result of entropy accumulation in the quantum critical regime~\cite{Garst2005}.

Upon applying this concept to {\rucl} two remarks are in order: (i) Its phase transition(s) can be driven by \emph{both} field and pressure, therefore both the field and pressure derivatives of the entropy will display sign changes, making $\Gamma$ a suitable probe to detect QCPs. (ii) For the qualitative analysis, we use the linear $c^*$-axis (instead of volume) TE coefficient $\alpha_{c^*}$, which is much larger compared to that along the other directions \cite{He2018}.

\begin{figure}
\includegraphics[scale=0.9]{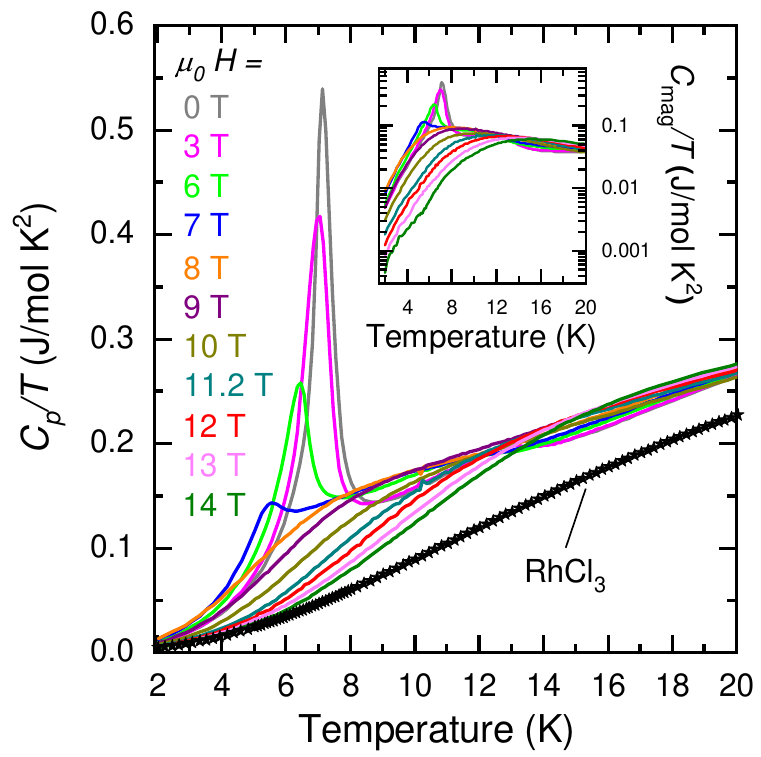}
\caption{\label{Fig_cp}
Specific-heat coefficient $C_p/T$ of {\rucl} as function of temperature for applied magnetic fields $\vec H \parallel$ Ru-Ru bonds, together with $C_p/T$ data for nonmagnetic RhCl$_3$ (black stars). The inset shows the magnetic contribution $C_{p,\rm mag}/T$ of {\rucl} on a semi-logarithmic scale.}
\end{figure}

To calculate the Gr\"uneisen ratio, the specific heat of {\rucl} is needed. The corresponding specific-heat coefficient $C_p/T$ is shown in Fig.~\ref{Fig_cp} as a function of temperature and magnetic field $\vec H \parallel$ Ru-Ru bonds. The data agree with previous reports for unknown in-plane field directions~\cite{Wolter2017,Loidl2019}, with a single magnetic transition at zero field at $T_N = 7.1(1)$\,K defined at the peak position of $C_p/T$. For $\mu_\mathrm{0}H \gtrsim 8$\,T the peak and thus the magnetic long-range order disappears, as expected.

For both $\alpha_{c^*}$ and $C_p$ the phononic contribution is assumed to be field-independent and had to be determined and subtracted from the {\rucl} data.
The phononic contribution to the specific heat of {\rucl} was approximated by the specific heat of the non-magnetic structural analog compound RhCl$_3$ (see Fig.~\ref{Fig_cp}), after scaling its experimental specific heat curve by the Lindemann factor~\cite{Lindemann1910}, which was found to be 1.000059.
For the phononic contribution of $\alpha_{c^*}(T)$ we used the {\rucl} data at $7.75$\,T as an approximation. Alternative schemes are discussed in Appendix~\ref{app:phonon}.

\begin{figure}[tb]
\includegraphics[scale=0.9]{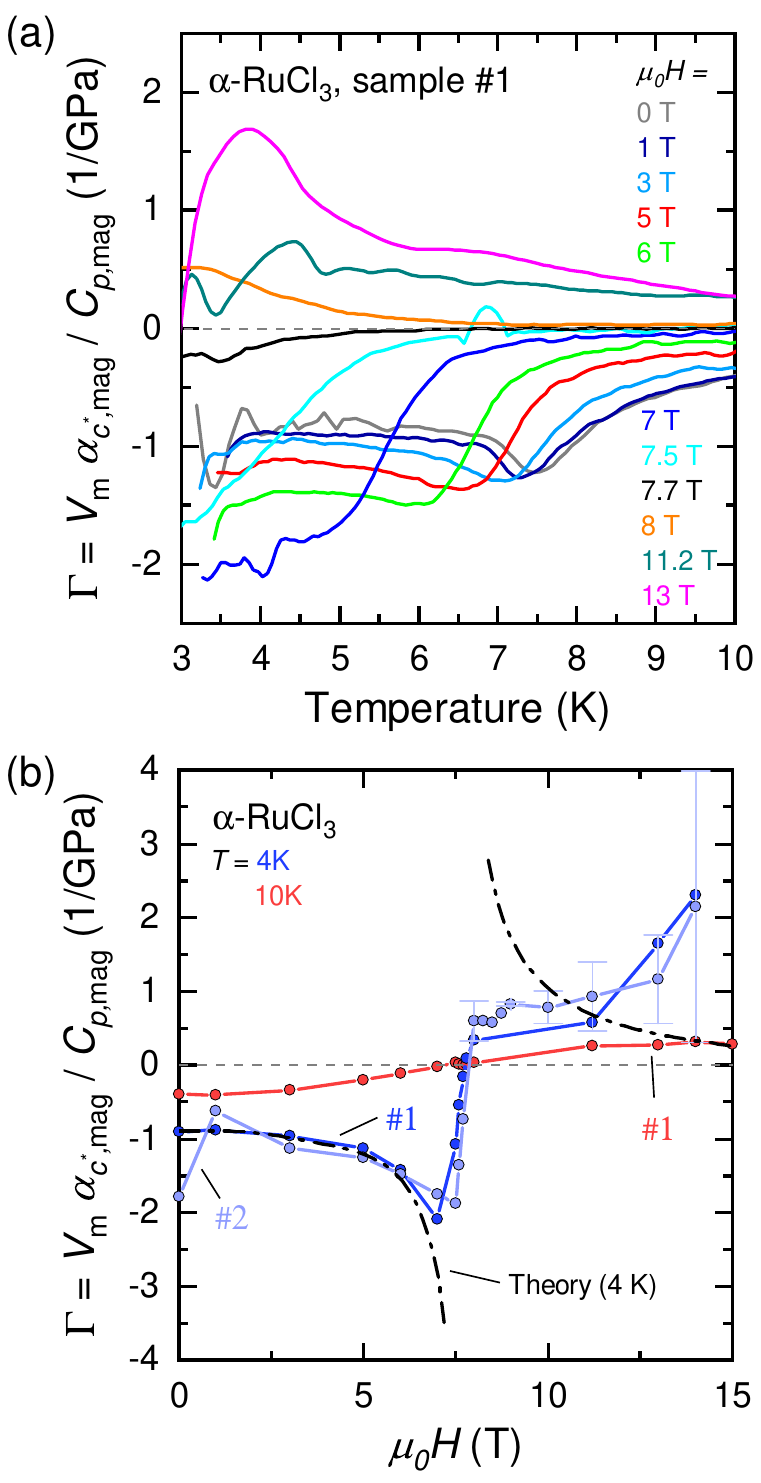}
\caption{\label{gamma}
The Gr\"uneisen ratio $\Gamma$ of {\rucl} in a magnetic field along the Ru-Ru bonds. (a) $\Gamma(T)$ for different fields up to 13\,T. (b) $\Gamma(H)$ for $4$\,K and $10$\,K and for two different samples (\#1, \#2). The dash-dotted line shows the theoretical calculation for $\Gamma$ at $4$\,K whose overall amplitude has been scaled to match the low-field part of the data. Close to $H_{\mathrm c1}$ no results are shown as the spin-wave approximation becomes unreliable, see text.}
\end{figure}

The resulting Gr\"uneisen ratio $\Gamma$ is depicted in Figs.~\ref{gamma}(a) and \ref{gamma}(b) as function of temperature and applied magnetic field $\vec H \parallel$ Ru-Ru bonds, respectively. A more comprehensive data set, including measurements of different samples, is shown in Appendix~\ref{app:sample}.
As function of field, $\Gamma$ changes sign at $H_{\mathrm c1}$ as expected. For fields below $H_{\mathrm c1}$, $\Gamma(T)$ displays a peak at the N\'eel temperature $T_{\rm N}(H)$, while becoming small at high $T$, Fig.~\ref{gamma}(a). Moreover, below $H_{\mathrm c1}$ and at low $T$, $\Gamma(H)$  has its largest magnitude close to $H_{\mathrm c1}$, Fig.~\ref{gamma}(b). The low-field part thus appears consistent with quantum critical phenomenology~\cite{Garst2005}, and the fact that $\Gamma(T)$ does not change sign at a temperature $T\gtrsim T_\mathrm{N}$ (Fig.~\ref{gamma}(a)) implies a large fluctuation regime above the quasi-two-dimensional magnetic transition. Together, the data signifies a continuous quantum phase transition at $H_{\mathrm c1}$ -- the same conclusion was reached earlier based on a detailed analysis of low-T specific-heat measurements \cite{Wolter2017}.

The low-$T$ data for $\Gamma$ above $H_{\mathrm c1}$ are again anomalous, in that there is no appreciable field dependence in $\Gamma(H)$ between $8$ and $11$\,T. As a result, the behavior of $\Gamma(H)$ around $H_{\mathrm c1}$ is rather asymmetric, Fig.~\ref{gamma}(b). We note that $\Gamma$ for fields of $13$\,T and above displays large error bars at low $T$ because both $\alpha_{c^*,\rm mag}$ and $C_{p,\rm mag}$ become very small as the magnetic excitations are gapped out.

\subsection{Magnetostriction}

\begin{figure}
\includegraphics[scale=1.1]{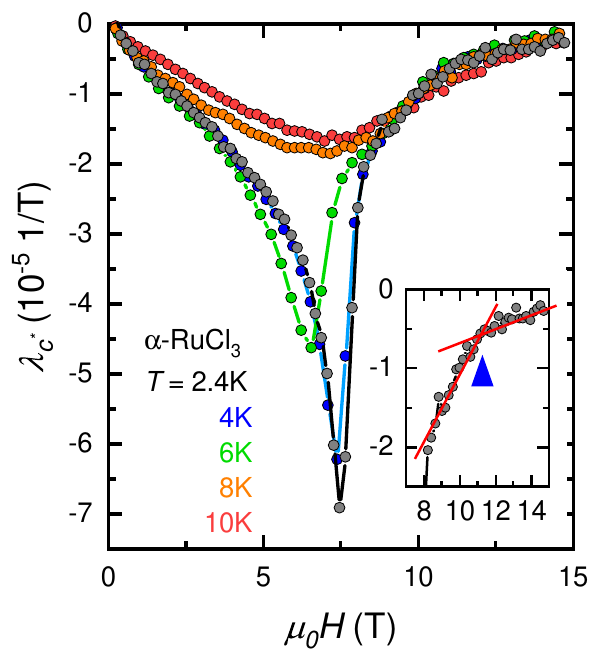}
\caption{\label{ms}
The linear MS coefficient $\lambda_{c^*}$ as function of field and different temperatures (sample 1). The inset highlights the kink observed in $\lambda_{c^*}$ at 2.4~K.}
\end{figure}

Field-driven phase transitions can be efficiently studied in MS experiments, measuring the length change as function of the applied field at constant $T$. Results for the linear MS coefficient along $c^*$,
\begin{equation}
\label{eq:lambda}
\lambda_{c^*} = \frac{\partial}{\partial (\mu_0 H)}
\frac{\Delta L_{c^*}(T,\mu_0H)}{L_{c^*} (300\,{\rm K},0\,{\rm T})},
\end{equation}
are displayed in Fig.~\ref{ms}. At $T=2.4$\,K the continuous transition at $H_{\mathrm c1}$ causes a sharp peak in $\lambda_{c^*}$ at $\mu_0H = 7.8(2)$\,T, which broadens and shifts to lower fields upon increasing temperature, thus tracking $T_\mathrm{N}(H)$. At $10$\,K (i.e., above $T_\mathrm{N}$), $\lambda_{c^*}$ deviates from a linear field dependence that would be expected for a usual paramagnet. This is again related to the large fluctuation regime reaching up to temperatures of the exchange couplings ($\sim 50$\,K); we recall that {\rucl} has been characterized as a ``Kitaev paramagnet'' in this regime~\cite{Do2017,Jansa2018}.

A striking feature is seen in the low-$T$ MS data above~$H_{\mathrm c1}$: While the 2.4\,K data of $\lambda_{c^*}(H)$ show no signature for a second continuous transition, the curve displays a clear kink at $\mu_0 H_{\mathrm c2} \approx 11$\,T, see inset of Fig.~\ref{ms}. Upon increasing the temperature, the kink position varies only weakly while the kink magnitude (i.e. the change in slope) decreases, with the kink disappearing for temperatures above $8$\,K. These observations suggest the existence of an additional low-temperature quantum regime whose field width remains finite at the lowest temperatures and should therefore be distinguished from the narrow quantum critical regime near $H_\mathrm{c1}$ and the semiclassical partially polarized regime above~$H_\mathrm{c2}$.

\section{Theory}
\subsection{Methods} \label{sec:th-methods}
\subsubsection{$J_1$-$K_1$-$\Gamma_1$-$J_3$ model}

For a theoretical description of the thermodynamic behavior of {\rucl}, we employ a minimal spin model containing nearest-neighbor Heisenberg $J_{1}$, Kitaev $K_{1}$, and off-diagonal $\Gamma_{1}$ interaction, as well as a third-nearest-neighbor Heisenberg $J_{3}$ interaction on the two-dimensional honeycomb lattice \cite{winter2016},
\begin{align}
\mathcal{H} & =\sum_{\langle ij\rangle}\left[J_{1}\vec{S}_{i}\cdot\vec{S}_{j}+K_{1}S_{i}^{\gamma}S_{j}^{\gamma}+\Gamma_{1}(S_{i}^{\alpha}S_{j}^{\beta}+S_{i}^{\beta}S_{j}^{\alpha})\right]
\nonumber \\ & \quad
+\sum_{\lllangle ij\rrrangle}J_{3}\vec{S}_{i}\cdot\vec{S}_{j}.\label{eq:jkg}
\end{align}
Here, $(\alpha,\beta,\gamma)=(x,y,z)$ on a nearest-neighbor $z$ bond, for example. The spin quantization axes point along the cubic axes of the RuCl$_{6}$ octahedra, such that the $[111]$ direction is perpendicular to the honeycomb $ab$ plane (referred to as $c^{*}$ axis) and the in-plane $[\bar{1}10]$ direction points along a Ru-Ru nearest-neighbor bond of the honeycomb lattice. The model displays a $C_3^\ast$ symmetry of combined threefold rotations in real and spin space; a possible trigonal distortion spoiling this symmetry is neglected. Additional off-diagonal couplings, dubbed $\Gamma'$, are symmetry-allowed but assumed to be negligible.

We are interested in the behavior of this model in the presence of an external magnetic field, described by the Hamiltonian
\begin{align}
\mathcal{H}'=\mathcal{H}-g\mu_{0}\mu_{\mathrm{B}}\sum_{i}\vec{H}\cdot\vec{S}_{i}
\end{align}
with $\vec{H}\parallel[\bar{1}10] \parallel$ Ru-Ru bonds.
Here, $g\mu_{\mathrm{B}}\vec{S}$ corresponds to the effective moment of the $J_{\text{eff}}=1/2$ states in the crystal, $g \equiv g_{ab}$ is the in-plane $g$ factor, and $\mu_\mathrm B$ the Bohr magneton.

\subsubsection{Thermodynamic relations}

We are interested in calculating changes of the sample length perpendicular to the $ab$ plane. Given the anisotropy of the {\rucl} crystal and the high sensitivity of the magnetic couplings to its structure, it is therefore important to also distinguish uniaxial from hydrostatic pressure.
We begin by writing down the differential of the Helmholtz free energy
\begin{equation}
\mathrm dF=-S \mathrm dT+\int\!\mathrm d^{3}r\,\sigma_{ij} \mathrm d \eta_{ij}-g\mu_{0}\mu_{\mathrm{B}}M \mathrm d H,\label{eq:dF}
\end{equation}
where $S$ denotes the entropy, $\sigma_{ij}$ and $\eta_{ij}$ are respectively the stress and strain tensors, $H = |\vec H|$ is the field strength, and $g \mu_\mathrm{B} M = g \mu_\mathrm{B} \sum_i |\langle \vec S_i \rangle|$ corresponds to the uniform magnetization, which we assume to be parallel to the magnetic field~\cite{janssen2017}. The spatial integral goes over the volume of the undeformed crystal \cite{landau1959,chaikin2000}.

Now, we may refine our description by taking the $C_{3}^\ast$ symmetry of our model into account. Indeed, this property implies that, under homogeneous stress, the system has only two independent length changes, namely of $\ell_{c^*}$ along the $c^{*}$ axis and $\ell_{ab}$ perpendicular to it. Furthermore, it guarantees that $\eta_{ij}$ becomes diagonal in a coordinate system which has one of its axes parallel to $c^{*}$. Thus, if we assume that stress is homogeneous throughout the sample, we may rewrite Eq.~\eqref{eq:dF} in the form
\begin{equation}
\mathrm dF=-S \mathrm dT+V\sigma_{i}\mathrm d\eta_{i}-g\mu_{0}\mu_{\mathrm{B}}M \mathrm dH\label{eq:dF homogeneous},
\end{equation}
where $V$ denotes the volume of the undeformed crystal, and we employ the shorthand notations $\eta_{i}\equiv\eta_{ii}$ and $\sigma_{i}\equiv\sigma_{ii}$. Each diagonal strain element $\eta_{i}$ then encodes information about the elongation along the $i$-th principal axis, such that $\mathrm d\eta_{i}=\mathrm d\ln\ell_{i}$
\cite{barrera2005,landau1959}.
In the following, we focus on a situation with uniaxial stress $\sigma_{c^*} \equiv - p_{c^*}$ along the $c^*$ axis, as relevant for the experiment~\cite{signnote}.

The Maxwell relations then read
\begin{align}
\alpha_{c^*} & =\left(\frac{\partial\ln\ell_{c^*}}{\partial T}\right)_{\sigma,H}=\frac{1}{V}\left(\frac{\partial S}{\partial\sigma_{c^*}}\right)_{\sigma_{ab},T,H},\label{eq:alpha def}\\
\lambda_{c^*} & =\left[\frac{\partial\ln\ell_{c^*}}{\partial\left(\mu_{0}H\right)}\right]_{\sigma,T}=\frac{\mu_{\mathrm{B}}}{V}\left[\frac{\partial\left(gM\right)}{\partial\sigma_{c^*}}\right]_{\sigma_{ab},T,H}, \label{eq:lambda def}
\end{align}
where the $\sigma_{ab}$ on the right-hand sides of Eqs.~\eqref{eq:alpha def} and \eqref{eq:lambda def} serve as a reminder that all stresses but $\sigma_{c^*}$ are to be held constant in carrying out the derivatives.

Within our two-dimensional model, it is impossible to compute the observables directly from Eqs.~\eqref{eq:alpha def} and \eqref{eq:lambda def}. Instead, we must consider how uniaxial stress along $c^\ast$ affects the microscopic parameters $\mathcal{J} \in \{J_{1},K_{1},\Gamma_{1},J_{3},g\}$ contained in the Hamiltonian, so that
\begin{align}
\alpha_{c^*} & = \frac{1}{V}\sum_{\mathcal{J}}\frac{\partial S}{\partial\mathcal{J}}\frac{\partial\mathcal{J}}{\partial\sigma_{c^*}},
	&
\lambda_{c^*} & = \frac{\mu_{\mathrm{B}}}{V}\sum_{\mathcal{J}}\frac{\partial\left(gM\right)}{\partial\mathcal{J}}\frac{\partial\mathcal{J}}{\partial\sigma_{c^*}}.\label{eq:alpha lambda micro}
\end{align}

\subsubsection{Pressure dependence of model parameters}

As the expressions in Eq.~\eqref{eq:alpha lambda micro} illustrate, the sensitivity of each microscopic parameter to stress plays a key role in determining $\alpha_{c^*}$ and $\lambda_{c^*}$. For small distortions we may expand $\mathcal J$ up to first order in $\sigma_{c^*}$
\begin{equation}
\mathcal{J}\left(\sigma_{c^*}\right)\approx\mathcal{J}_{0}\left[1+n_{\mathcal{J}}\left(\sigma_{c^*}-\sigma_{0}\right)\right],
\label{eq:J(pc) expansion}
\end{equation}
where $\sigma_{0}$ represents ambient stress and $\mathcal J_0$ is the corresponding unperturbed value of the model parameter, and we have defined the expansion coefficient
\begin{equation}
n_{\mathcal{J}}:=\frac{1}{\mathcal{J}_{0}}\left.\frac{\partial\mathcal{J}}{\partial\sigma_{c^*}}\right|_{\sigma_{c^*}=\sigma_{0}}.
\label{eq:nJ def}
\end{equation}
A positive $n_{\mathcal{J}}$ therefore means that the absolute value of $\mathcal{J}$ increases in response to increasing tensile stress $\sigma_{c^*}$ (i.e., decreasing uniaxial pressure along $c^*$).

As the exchange couplings in {\rucl} sensitively depend on bond lengths and angles~\cite{Yadav2016, winter2016}, the pressure dependence of the model parameters is not easily modeled. Comprehensive \emph{ab initio} information is presently lacking (see, however, Ref.~\onlinecite{yadav2019}). We therefore treat the various $n_{\mathcal{J}}$ as free parameters, aiming at reproducing the main features of the experimental results in regimes where spin-wave theory is reliable.

\subsubsection{Spin-wave theory}

We now compute the Helmholtz free energy on the level of linear spin-wave theory (LSWT). This semiclassical approach is expected to provide reliable results at low temperatures in both the ordered and the polarized high-field phases where the number of magnon excitations is small. It is, however, not reliable (i) at low fields for temperatures comparable to or above the N\'eel temperature $T_\mathrm{N}$ and (ii) at high fields for temperatures above the spin gap. This includes the quantum critical regime near $H_{\mathrm c1}$.

At zero field, the Hamiltonian \eqref{eq:jkg} hosts three different zigzag patterns as degenerate classical ground states. However, such a degeneracy is lifted by a $[\bar{1}10]$ field, which selects the configuration with zigzag chains running perpendicularly to it. This happens because the corresponding zero-field order is normal to the $[\bar{1}10]$ axis, so that the spins cant uniformly in response to the magnetic field \cite{janssen2017}. Canting increases until the critical field, $H_{\mathrm c1}$, is reached and a continuous transition from the canted zigzag to a partially polarized state takes place.

We employ a standard procedure involving the Holstein-Primakoff transformation, whereby quantum fluctuations with respect to the classical ground state are described as magnonic excitations \cite{holstein1940,blaizot1986,rau2018}; see Ref.~\onlinecite{janssen2019} for a pedagogic introduction in the context of Heisenberg-Kitaev models.
This amounts to introducing magnon creation and annihilation operators $a_{i\nu}^{\dagger}$ and $a_{i\nu}$ at site $\nu$ in the magnetic unit cell $i$. The index $\nu$ runs from 1 to the number of magnetic sublattices, $N_{s}$, so that $N_{s}=2$ and 4 in the polarized and canted zigzag phases, respectively. To the leading nontrivial order, we arrive at a quadratic spin-wave Hamiltonian in momentum space,
\begin{equation}
\mathcal{H}_{\mathrm{LSW}}=N\epsilon_\mathrm{cl}+\frac{1}{2}\sum_{\textbf{k}}\left(\Psi_{\textbf{k}}^{\dagger}\mathbb{M}_{\textbf{k}}\Psi_{\textbf{k}}-\text{Tr}\mathbb{A}_{\textbf{k}}\right),\label{eq:LSW hamiltonian}
\end{equation}
where $N$ is the total number of sites, $\epsilon_\mathrm{cl}$ is the classical energy per site, $\Psi_{\textbf{k}}^{\dagger}=\left(a_{\textbf{k}1}^{\dagger}\ldots a_{\textbf{k}N_{s}}^{\dagger},a_{-\textbf{k}1}\ldots a_{-\textbf{k}N_{s}}\right)$, and the summation is over all momenta in the magnetic Brillouin zone.
The $2N_{s}\times2N_{s}$ matrix $\mathbb{M}_{\textbf{k}}$
can be written in terms of two $N_{s}\times N_{s}$ submatrices, $\mathbb{A}_{\textbf{k}}$ and $\mathbb{B}_{\textbf{k}}$, as
\begin{equation}
\mathbb{M}_{\textbf{k}}=\begin{pmatrix}\mathbb{A}_{\textbf{k}} & \mathbb{B}_{\textbf{k}}\\
\mathbb{B}_{\textbf{k}}^{\dagger} & \mathbb{A}_{-\textbf{k}}^{\text{T}}
\end{pmatrix}.
\end{equation}
$\mathcal{H}_{\mathrm{LSW}}$ can then be diagonalized via
a Bogoliubov transformation \cite{janssen2019}, from which we obtain the eigenenergies~$\epsilon_{\textbf{k}\nu} > 0$. This step was performed analytically in the partially polarized phase, but required a numerical approach in the zigzag ordered phase.

In these terms, we may write the Helmholtz free energy for a system of non-interacting bosons as
\begin{equation}
F=E_\mathrm{gs}+\beta^{-1} \sum_{\textbf{k}\nu}\ln\left(1-e^{-\beta\epsilon_{\textbf{k}\nu}}\right),\label{eq:free energy}
\end{equation}
where $\beta=1/(k_{\mathrm{B}}T)$ is the inverse temperature and
\begin{equation}
E_\mathrm{gs}=N\epsilon_\mathrm{cl}+\frac{1}{2}\sum_{\textbf{k}}\left(\sum_{\nu=1}^{N_{s}}\epsilon_{\textbf{k}\nu}-\text{Tr}\,\mathbb{A}_{\textbf{k}}\right)\label{eq:Egs}
\end{equation}
denotes the ground-state energy including the leading-order quantum corrections. Then, by combining Eqs.~\eqref{eq:alpha lambda micro} and \eqref{eq:free energy}, we find
\begin{equation}
\alpha_{c^*}=-\frac{k_{\mathrm B} \beta^2}{V} \sum_{\mathcal{J}}\mathcal{J}_{0}n_{\mathcal{J}}\sum_{\textbf{k}\nu}\frac{e^{\beta\epsilon_{\textbf{k}\nu}}\epsilon_{\textbf{k}\nu}}{\left(e^{\beta\epsilon_{\textbf{k}\nu}}-1\right)^{2}}\frac{\partial\epsilon_{\textbf{k}\nu}}{\partial\mathcal{J}}\label{eq:alpha}
\end{equation}
and
\begin{align}
\lambda_{c^*} & =
-\sum_{\mathcal{J}}\frac{\mathcal{J}_{0}n_{\mathcal{J}}}{\mu_{0}V}
\Biggl\{ N\frac{\partial^{2}}{\partial\mathcal{J}\partial H}\left(\epsilon_\mathrm{cl}-\frac{\text{Tr\,}\mathbb{A}_{\textbf{k}}}{2N_{s}}\right)
\nonumber \\ & \quad
+\sum_{\textbf{k}\nu}\biggl[\left(\frac{1}{2}+\frac{1}{e^{\beta\epsilon_{\textbf{k}\nu}}-1}\right)\frac{\partial^{2}\epsilon_{\textbf{k}\nu}}{\partial\mathcal{J}\partial H}
\nonumber \\ & \quad
-\frac{\beta e^{\beta\epsilon_{\textbf{k}\nu}}}{\left(e^{\beta\epsilon_{\textbf{k}\nu}}-1\right)^{2}}\frac{\partial\epsilon_{\textbf{k}\nu}}{\partial H}\frac{\partial\epsilon_{\textbf{k}\nu}}{\partial\mathcal{J}}\biggr]\Biggr\}.
\label{eq:lambda}
\end{align}
Note that, in order to arrive at Eq.~\eqref{eq:lambda}, we have used the fact that $\text{Tr\,}\mathbb{A}_{\textbf{k}}$ is momentum-independent for the phases we consider here. Another noteworthy point is that both $\alpha_{c^*}$ and $\lambda_{c^*}$ are given as a linear combination of terms which result from varying one microscopic parameter at a time. While this does not hold beyond the approximation \eqref{eq:J(pc) expansion}, it allowed us to analyze an arbitrarily wide range of sets $\left\{ n_{\mathcal{J}}\right\} $ without demanding extra computational time.

A third quantity of interest is the magnetic heat capacity at constant strain $\eta$,
\begin{equation}
C_{\mathrm{\eta}}\left(T,H\right)=k_{\mathrm{B}} \beta^2 \sum_{\textbf{k}\nu}\frac{e^{\beta\epsilon_{\textbf{k}\nu}}\epsilon_{\textbf{k}\nu}^{2}}{\left(e^{\beta\epsilon_{\textbf{k}\nu}}-1\right)^{2}}\;.\label{eq:CV}
\end{equation}
In the following, we will compare $\alpha_{c^*}/C_{\eta}$ to the Gr\"uneisen ratio measured in the experiments, even though both observables are not strictly equal. That is because the Gr\"uneisen ratio is defined as the ratio between the TE coefficient and the heat capacity at constant stress rather than constant strain.

\subsubsection{Parameter sets}

The values for the exchange couplings in Eq.~\eqref{eq:jkg} can be estimated from \textit{ab initio} calculations~\cite{winter2016}; however, we find better agreement with our experimental data by using a slightly adapted parameter set that has recently been suggested by comparing with zero-field neutron scattering data~\cite{winter2017}:
\begin{align}
(J_{1},K_{1},\Gamma_{1},J_{3}) & =(-0.1,-1.0,+0.5,+0.1) A,\label{eq:couplings}
\end{align}
where $A$ is a global energy scale, which we adjust to $A = 4.31$\,meV, so that the continuous classical transition between the canted zigzag and the polarized phase occurs at  $\mu_\mathrm{0}H_{\mathrm c1}=7.8$\,T for $g=2.8$ and in-plane fields $\vec H \parallel [\bar110]$, as in the experiment~\cite{Majumder2015}.
The above model has previously been shown to reproduce features found in a number of experiments on {\rucl} \cite{winter2017, janssen2017, winter2018, Wolter2017, kelley2018b, winter2017review, janssen2019}, although it might require adjustments when finite interlayer interactions \cite{Balz2019} are taken into account \cite{janssen2020}.

In addition to the values of the model parameters at ambient pressure, we require their pressure dependence. In order to work with dimensionless parameters, we have chosen to normalize the $n_{\mathcal{J}}$ with respect to $n_{\Gamma_1}>0$. The latter represents an overall fitting factor that can be determined by matching the global scale of the experimental data. A good fit is obtained for $n_{\Gamma_1} = 0.9$\,GPa$^{-1}$, cf.\ Fig.~\ref{gamma}(b).

As a first trial, we looked into a scenario where uniaxial pressure would affect all exchange couplings, but not the $g$ factor. Then, we adjusted the remaining free parameters to reproduce as many qualitative features from the experimental data as possible, regarding both the TE and the MS measurements. This approach lead us to the set of coefficients labeled by Set 1 in Table \ref{tab:parameter sets}.

However, as we shall explain in the following, our MS results motivated us to consider a second scenario, whereby we removed the constraint $n_{g}=0$. In this way, a new comparison to the experimental data yielded the set we will refer to as Set 2.

\begin{table}[tb]
\renewcommand{\arraystretch}{1.2}
\caption{Different sets of expansion coefficients $n_{\mathcal{J}}$ describing the stress dependence of the model parameters, see Eq.~\eqref{eq:nJ def}. The value of $n_{\Gamma_1}$~is used as reference scale, see text.\label{tab:parameter sets}}
\begin{tabular*}{\linewidth}{@{\extracolsep{\fill}} l c c c c}
\hline\hline
 & $n_{J_{1}}/n_{\Gamma_1}$ & $n_{K_{1}}/n_{\Gamma_1}$ & $n_{J_{3}}/n_{\Gamma_1}$ & $n_{g}/n_{\Gamma_1}$ \tabularnewline
\hline
Set 1 & 0.3 & 0.3 & 1.6 & 0 \tabularnewline
Set 2 & 0.5 & 0.75 & 0.56 & -0.65 \tabularnewline
\hline\hline
\end{tabular*}
\end{table}

Before proceeding, we note that both Set 1 and Set~2 predict that the magnitudes of all exchange couplings decrease with increasing pressure along the $c^{*}$ axis. Although the intricate nature of the quantum chemistry behind the model prevents a clear judgment on how reasonable such an observation is, a previous \emph{ab initio} study has suggested the same trend after considering a similar model for {\rucl} \cite{yadav2019}.

\subsection{Thermal expansion}

\begin{figure*}[tbp]
\includegraphics[width=\linewidth]{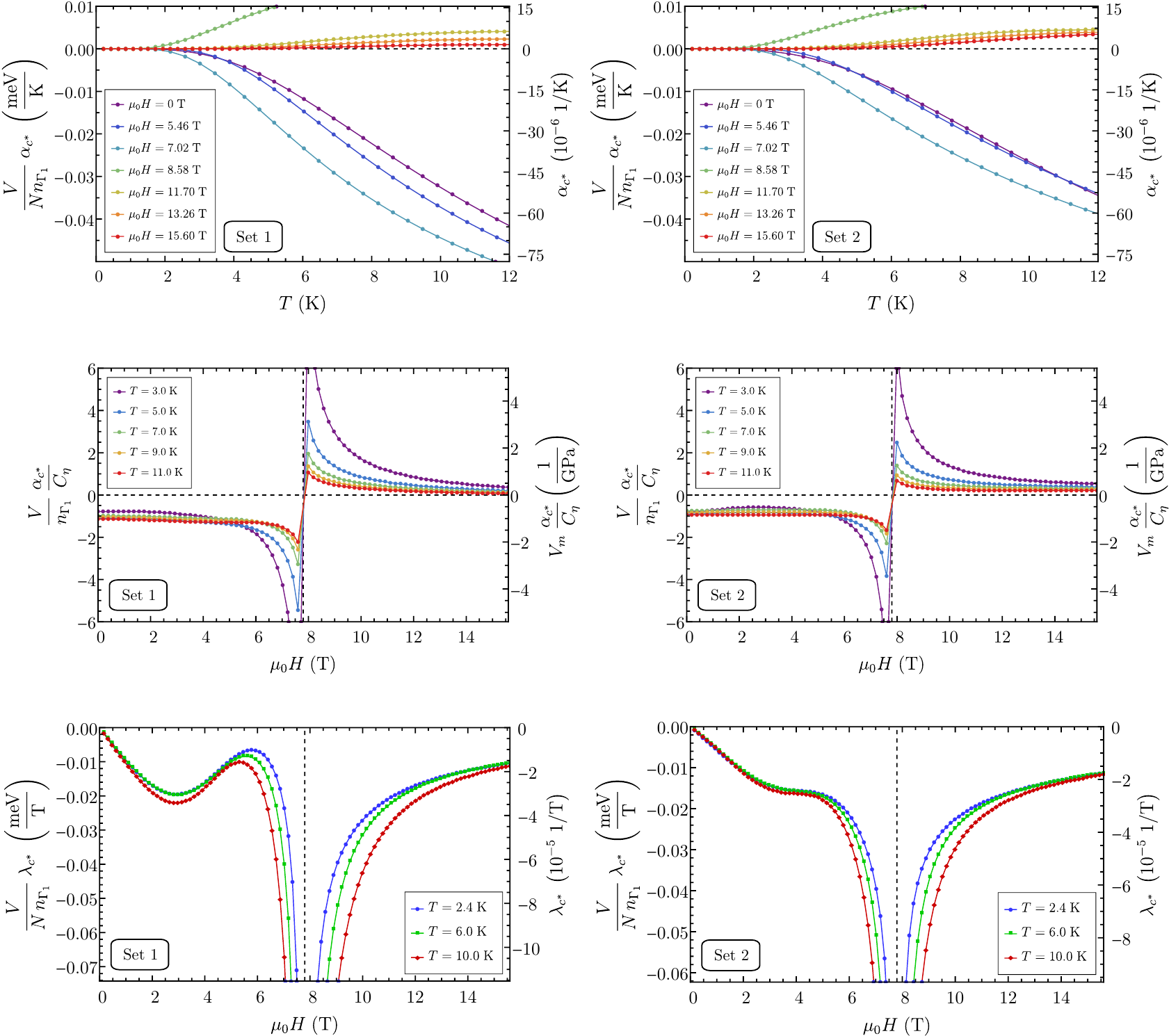}
\caption{%
From top to bottom: TE coefficient, Gr\"uneisen ratio and MS plots for the two sets of microscopic parameters given in Table \ref{tab:parameter sets}. In the plots where the independent variable is the magnetic field, the vertical gridline represents the critical field, $\mu_\mathrm{0}H_{\mathrm c1}=7.8$\,T. In all panels, the left vertical axis gives the theory result normalized to $n_{\Gamma_1}$, the right vertical axis shows experimental units using $n_{\Gamma_1} = 0.9$\,GPa$^{-1}$ and $V/N \simeq 92.8\,\text{\AA}^3$, corresponding to a molar volume $V_\text{m} \simeq 55.9\,\text{cm}^3/\text{mol}$~\cite{Johnson2015,Park2016,Cao2016}.
Our semiclassical analysis does not show any of the features interpreted as signatures of an additional regime at intermediate fields above $H_{\mathrm c1}$ and thus supports its genuine quantum character.
}
\label{fig:plots}
\end{figure*}

We now discuss our TE results, which are shown in the top row of Fig. \ref{fig:plots}. Both sets of coefficients presented in Table \ref{tab:parameter sets} reproduce gross features of the experimental data, such as: (i) $\alpha_{c^*}$ is negative for $H<H_{\mathrm c1}$ and positive for $H>H_{\mathrm c1}$; (ii) the magnitude of $\alpha_{c^*}$ is markedly smaller for $\mu_\mathrm{0}H\gtrsim11$\,T than for $\mu_\mathrm{0}H\lesssim7$\,T; (iii) $\alpha_{c^*}$ is suppressed by increasing $H$ at sufficiently high fields. Moreover, Set 1 also leads to the correct field trend in the zigzag phase, whereby the magnitude of $\alpha_{c^*}$ becomes larger as one increases $H$ at a fixed temperature. On the other hand, Set 2 only does so approximately, since the trend is spoiled near zero field.
We recall that LSWT does not capture the thermal phase transition at $T_\mathrm{N}$, hence a corresponding peak in $\alpha_{c^*}(T)$ is missing, and a comparison of experiment and theory for low fields should be restricted to $T<T_\mathrm{N}$.

Upon analyzing different combinations of the coefficients $n_{\mathcal{J}}$, we were unable to obtain a reasonable agreement with experiment without considering a comparatively large $n_{J_{3}}>0$. This interesting observation suggests that uniaxial pressure along the $c^{*}$ axis destabilizes the zigzag phase, since this particular type of magnetic ordering is favored by a positive $J_{3}$.

As an additional remark on the role of the expansion coefficients~$n_{\mathcal{J}}$, we point out that the correct field evolution for $H<H_{\mathrm c1}$ requires a sizable $n_{\Gamma_{1}}>0$. Increasing $n_{K_{1}}$ and $n_{J_{1}}$ tends to cancel this trend and even produce negative values for $\alpha_{c^*}$ in the polarized phase at temperatures larger than $15$\,K.

With that said, we emphasize that neither Set 1 and Set 2, nor any other combination of coefficients we considered in our analysis produces the non-monotonic behavior for $\alpha_{c^*}(T)$ at fields in the region between $H_{\mathrm c1}$ and $H_{\mathrm c2}$, as reported in experiment. Instead, we generally find that $\alpha_{c^*}$ increases monotonically at a fixed temperature as $H\rightarrow H_{\mathrm c1}^{+}$.

\subsection{Gr\"uneisen ratio}

Next, we discuss the evolution of the Gr\"uneisen ratio as a function of the magnetic field (see the second row of plots in Fig. \ref{fig:plots}). Our results show the correct signs above and below $H_{\mathrm c1}$, as we have enforced this by a careful analysis of the TE data.

Very close to $H_{\mathrm c1}$ magnon excitations proliferate and the non-interacting boson picture underlying LSWT becomes inadequate. Hence, the evolution of $\alpha_{c^*}$ and $\alpha_{c^*}/C_\eta$ through the critical field cannot be reliably computed within our approach. On general grounds \cite{Garst2005} we expect that both quantities evolve smoothly at fixed finite $T$ as function of $H$, with the exception of a singularity at $T_\mathrm{N}(H)$, crossing zero near $H_{\mathrm c1}$.

As noted above, however, we expect our calculations to yield correct results at sufficiently high fields and low temperatures, where the magnon excitation gap is comparable to or larger than $k_{\mathrm{B}}T$. A sample result of the theory for $T = 4$\,K in comparison with the experimental data, displayed in Fig.~\ref{gamma}(b), shows the field dependence of the calculated Gr\"uneisen ratio $\Gamma$ at a fixed low $T$. For a standard direct transition between zigzag and partially polarized phases, the semiclassical result should agree for all fields except very close to the quantum critical point at $H_{\mathrm c1}$. The match at low fields is convincing, however, the anomaly of the measured $\Gamma$ in the region above $H_{\mathrm c1}$ does not appear compatible with an interpretation in terms of spin-wave theory, suggesting the presence of a genuine quantum regime.

\subsection{Magnetostriction}

When we move to the results on the MS (bottom row of Fig. \ref{fig:plots}), we see that both sets of coefficients correctly produce negative values of $\lambda_{c^*}$ for the whole range of magnetic fields considered here. However, Set 1 notably leads to large non-monotonic variations around an inflection point in the zigzag phase which are not observed in experiment. The origin of this is in the behavior of the magnon spectrum which evolves in a highly non-trivial fashion with field.

When trying to reduce the intensity of this feature in $\lambda_{c^*}$, we verified that it becomes even larger if one, for instance, decreases $n_{K_{1}}$. In fact, without considering variations in $g$, we were unable to find a parameter set capable of smoothing out such a contortion while preserving the main characteristic of the TE coefficient. As far as we could check, this is only accomplished by taking $n_{g}<0$, which motivated us to consider the second set of coefficients.

We recall that LSWT does not produce critical behavior at $T_\mathrm{N}$, therefore $\lambda_{c^*}(H)$ displays a singularity at $H_{\mathrm c1}$ for all temperatures, instead of a singularity at $T_\mathrm{N}(H)$.

In regard to the behavior for $H>H_{\mathrm c1}$, our results do not bear any resemblance to the kink found in experiment. Together with the absence of a non-monotonic behavior in $\alpha_{c^*}$ around $8$\,T and the lack of the asymmetric, anomalous Gr\"uneisen ratio above $H_{\mathrm c1}$, this suggests that the physics in the regime between $H_{\mathrm c1}$ and $H_{\mathrm c2}$ cannot be fully accounted for semiclassically, in terms of a continuous field-induced opening of a spin gap alone. This supports the interpretation of our experimental data in terms of an exotic quantum regime in a finite low-temperature region above the quantum critical point at $H_{\mathrm c1}$.

\section{Discussion}

\begin{figure}[t]
\includegraphics[scale=1.05]{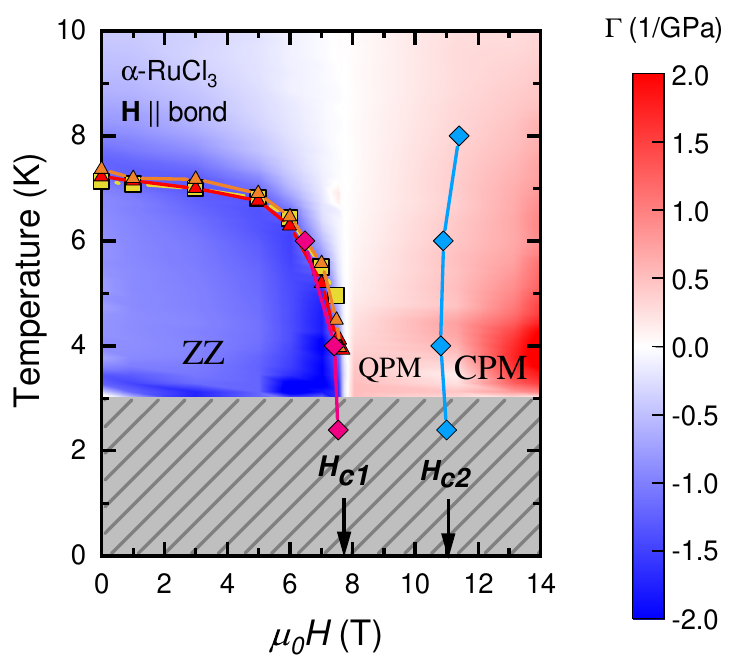}
\caption{\label{pd}
The $T$-$H$ phase diagram of {\rucl} as determined in this work for $\vec H \parallel$ Ru-Ru bonds ($C_p$: yellow squares, TE: red and orange triangles for samples \#1 and \#2, respectively, MS: pink and blue diamonds). Additionally, the Gr\"uneisen ratio $\Gamma (T,H)$ is shown in color scale (sample 1); these data are the same as in ~\ref{gamma} (a). The gray shaded region marks the limits for the color map. In addition to the zigzag (ZZ) ordered state at low fields and the conventional paramagnetic (CPM) state at high fields, our analysis suggests the presence of a third low-temperature regime that is of genuine quantum nature (QPM). The field-driven low-temperature transition at $H_{\mathrm c1}$ is continuous, whereas the one at $H_{\mathrm c2}$ is either a crossover or a weak first-order transition.}
\end{figure}

Based on these findings, we construct a temperature-field phase diagram of {\rucl} for $\vec H \parallel$ Ru-Ru bonds using our experimental thermal expansion, magnetostriction, and specific-heat results. It also includes the Gr\"uneisen ratio $\Gamma (T,H)$ in color scale, interpolated from the experimental data (see Sec.~\ref{subsec:exp-grueneisen}, Fig. ~\ref{gamma} (a)). The phase diagram, Fig.~\ref{pd}, shows three distinct low-temperature regimes: (i) the low-field phase with zigzag (ZZ) long-range order terminating at $\mu_0H_{\mathrm c1} = 7.8(2)$\,T, (ii) an intermediate quantum paramagnetic (QPM) regime between $\mu_0H_{\mathrm c1}$ and $\mu_0H_{\mathrm c2} \approx$ 11~T, and (iii) a conventional paramagnetic (CPM) state with a gapped magnon spectrum and partially polarized spins at high fields above $\mu_0H_{\mathrm c2}$.

We speculate that the QPM intermediate regime could possibly represent a topological quantum spin liquid as claimed by thermal Hall effect studies~\cite{Kasahara2018,Yokoi2020}. Such a phase is not symmetry-distinct from the CPM high-field phase and does not require it to be bounded by a thermal phase transition.
Our measurements of the Gr\"uneisen ratio did not show signs of quantum critical scaling near $H_{\mathrm c2}$, suggesting the absence of a second-order transition. The transition between the QPM and CPM regimes should therefore be either a crossover or a weak first-order transition. In the latter case, one should expect the transition line at $H_{\mathrm c2}$ to terminate at a critical endpoint at finite temperature.
We note that the signatures observed around $H_{\mathrm c2}$ may in principle also be related to a change in the character of the magnetic excitations as probed by Raman and THz spectroscopy, where indications for magnon bound states have been reported~\cite{Valenti2019,Wulferding2019}.
Near $H_{\mathrm c1}$, on the other hand, the Gr\"uneisen ratio exhibits characteristic quantum critical behavior, confirming the earlier proposal~\cite{Wolter2017, directionnote} of a quantum critical point at $\mu_0 H_{\mathrm c1} = 7.8(2)$ T and $T=0$.

\section{Conclusions}
Our high-resolution thermal expansion and magnetostriction measurements of {\rucl} along the $c^*$ axis confirm the field-induced suppression of long-range magnetic order at a critical field of $\mu_\mathrm{0}H_{\mathrm c1} = 7.8(2)$\,T, applied parallel to the Ru-Ru bonds, and provide thermodynamic evidence for quantum critical behavior at $H_{\mathrm c1}$ from an analysis of the Gr\"uneisen ratio.
A clear kink in the measured linear MS coefficient at $\mu_0 H_{\mathrm c2} \approx 11$\,T hints at an additional weak first-order phase transition, or a finite-temperature crossover, while an additional second-order phase transition above $H_{\mathrm c1}$ can be ruled out.
A comparison of our experimental data to calculations using a minimal lattice model, solved in the semiclassical limit via linear spin-wave theory, shows that the behavior at low fields appears well captured by semiclassical theory. In contrast, the regime between $H_{\mathrm c1}$ and $H_{\mathrm c2}$ is not explained by our minimal spin model.
While we cannot draw clear conclusions about the nature of the low-$T$ state at these intermediate fields, we speculate that this could possibly represent the topological quantum spin liquid suggested earlier~\cite{Kasahara2018,Yokoi2020}.

Our findings call for a more detailed experimental and theoretical study of the MS (and related quantities) in {\rucl} for different in- and out-of-plane field directions. Further theoretical work is needed to investigate possible field-driven transitions in and out of the putative spin liquid in Kitaev-based models~\cite{janssen2019} and trying to understand the field dependence of TE and MS in the spin-liquid phase.

\acknowledgements

We acknowledge insightful discussions with E. C. Andrade, G. Bastien, S. Biswat, K. Riedl, D. Kaib, S. Rachel, R. Valent{\'{\i}}, and S. M. Winter. This research has been supported by the Deutsche Forschungsgemeinschaft (DFG) through SFB 1143 (project id 247310070), the W\"urzburg-Dresden Cluster of Excellence on Complexity and Topology in Quantum Matter -- \textit{ct.qmat} (EXC 2147, project id 390858490), and the Emmy Noether program (JA2306/4-1, project id 411750675). P.M.C.\ was supported by the FAPESP (Brazil) Grant No.\ \mbox{2019/02099-0}. S.N.\ was supported by the Scientific User Facilities Division, Basic Energy Sciences, US DOE. D.G.M. acknowledges support from the Gordon and Betty Moore Foundation's EPiQS Initiative, Grant GBMF 9069.

\appendix

\section{Sample dependence}\label{app:sample}

\begin{figure}[b]
\includegraphics[scale=0.83]{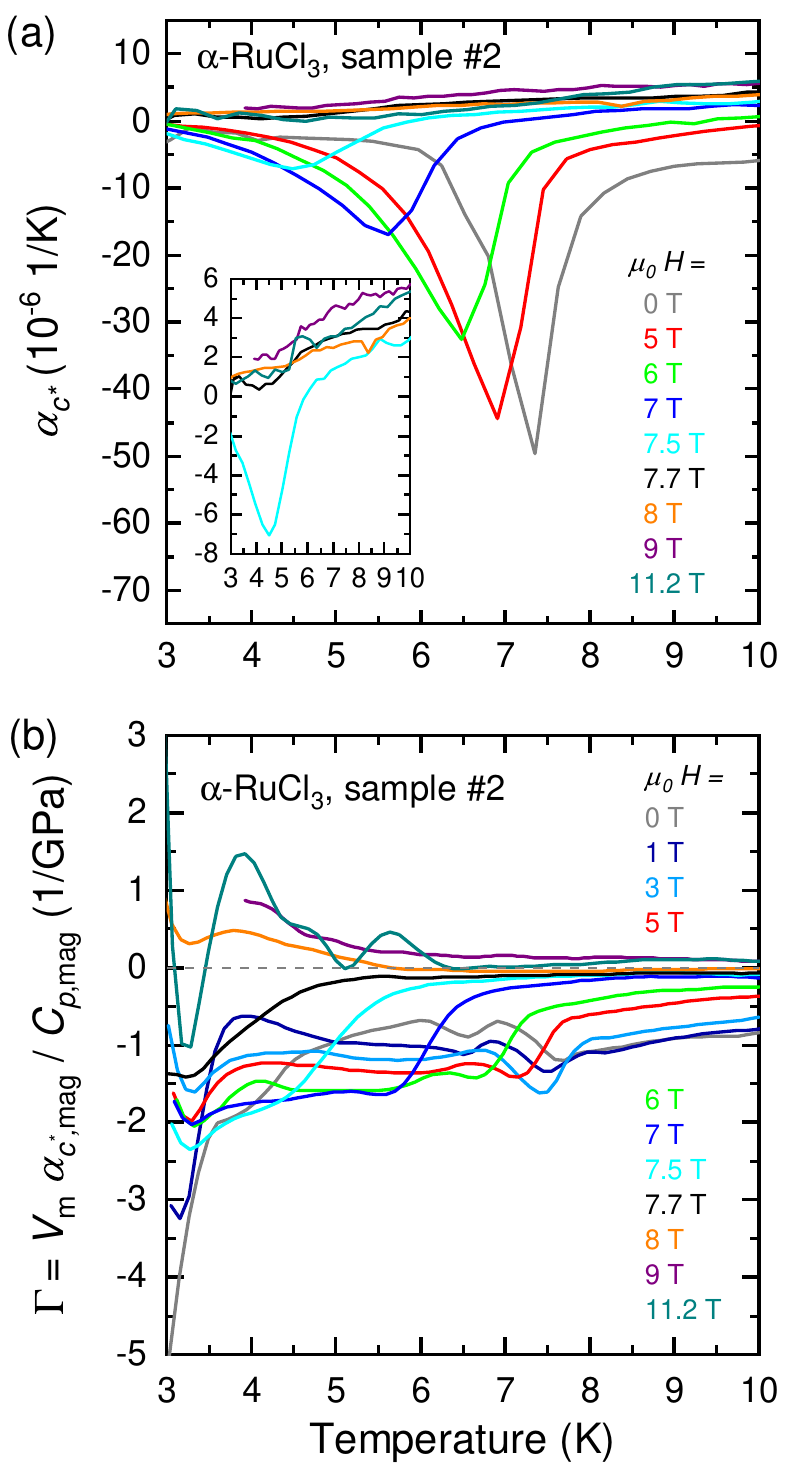}
\caption{\label{alpha_sample2}
(a) Temperature dependence of the linear TE coefficient perpendicular to the $ab$ plane, $\alpha_{c^*}$, for sample \#2 of {\rucl} for zero and applied magnetic fields $\vec H \parallel$ Ru-Ru bonds. The inset shows a zoom of the data for fields close to the critical field $\mu_0H_{\mathrm c1} = 7.8(2)$\,T.
(b) Gr\"uneisen ratio $\Gamma(T)$ of {\rucl} of sample \#2 parallel to the Ru-Ru bonds.}
\end{figure}

The TE was measured on two different samples of {\rucl} with similar crystal dimensions ($\sim$ 1 and $\sim$ 0.8~mm in thickness). In this way, sample dependencies due to crystal imperfections were tested. The linear TE coefficient $\alpha_{c^*}(T)$ and the corresponding Gr\"uneisen ratio $\Gamma(T)$ on sample \#2 are shown in Fig.~\ref{alpha_sample2}.
For comparison, the same observables obtained for sample \#1 are shown in Figs. \ref{alpha} and \ref{gamma}(a) above. The key features are identical, i.e., the sharp peak in $\alpha_{c^*}$ at the antiferromagnetic phase transition $T_\mathrm{N} = 7.2(1)$\,K, the sign change of $\alpha_{c^*}$ and $\Gamma(T)$ at the critical field $\mu_0H_{\mathrm c1} = 7.8(2)$\,T, and the quantum critical signature in $\Gamma(T)$.
The shallow maximum in $\alpha_{c^*}(T)$ at low temperatures and at $8$\,T, however, cannot be observed in sample \#2. This is probably related to the reduced thickness of sample \#2 leading to even smaller changes in the TE, being at the resolution limit of our experimental setup for these kind of thin 2D van-der-Waals materials.
It should be noted, however, that the linear TE coefficient at low temperatures and fields $\mu_0H \sim$ 9\,T is also slightly increased for sample \#2, see inset of Fig.~\ref{alpha_sample2}(a).

\begin{figure*}[p]
\centering
\includegraphics[scale=0.85]{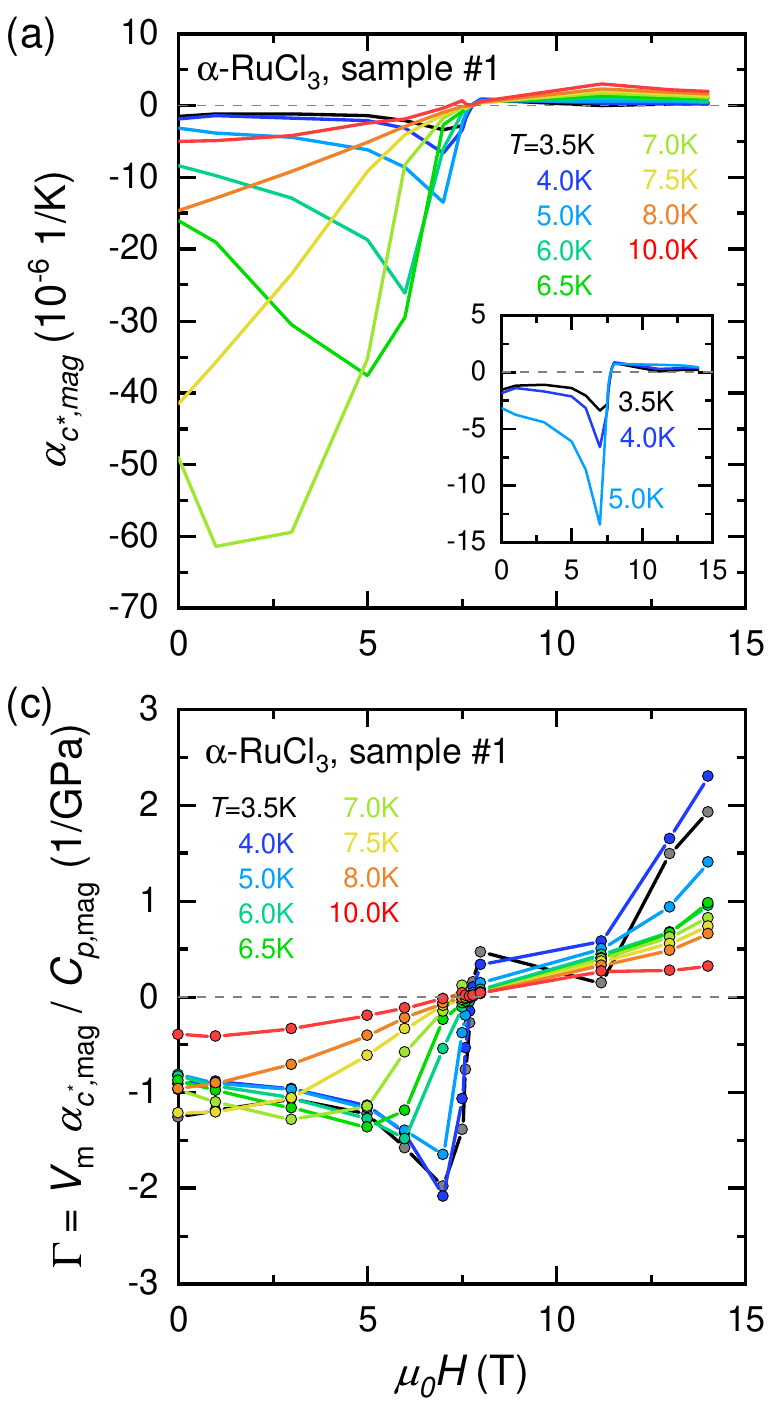}\hspace{2em}
\includegraphics[scale=0.85]{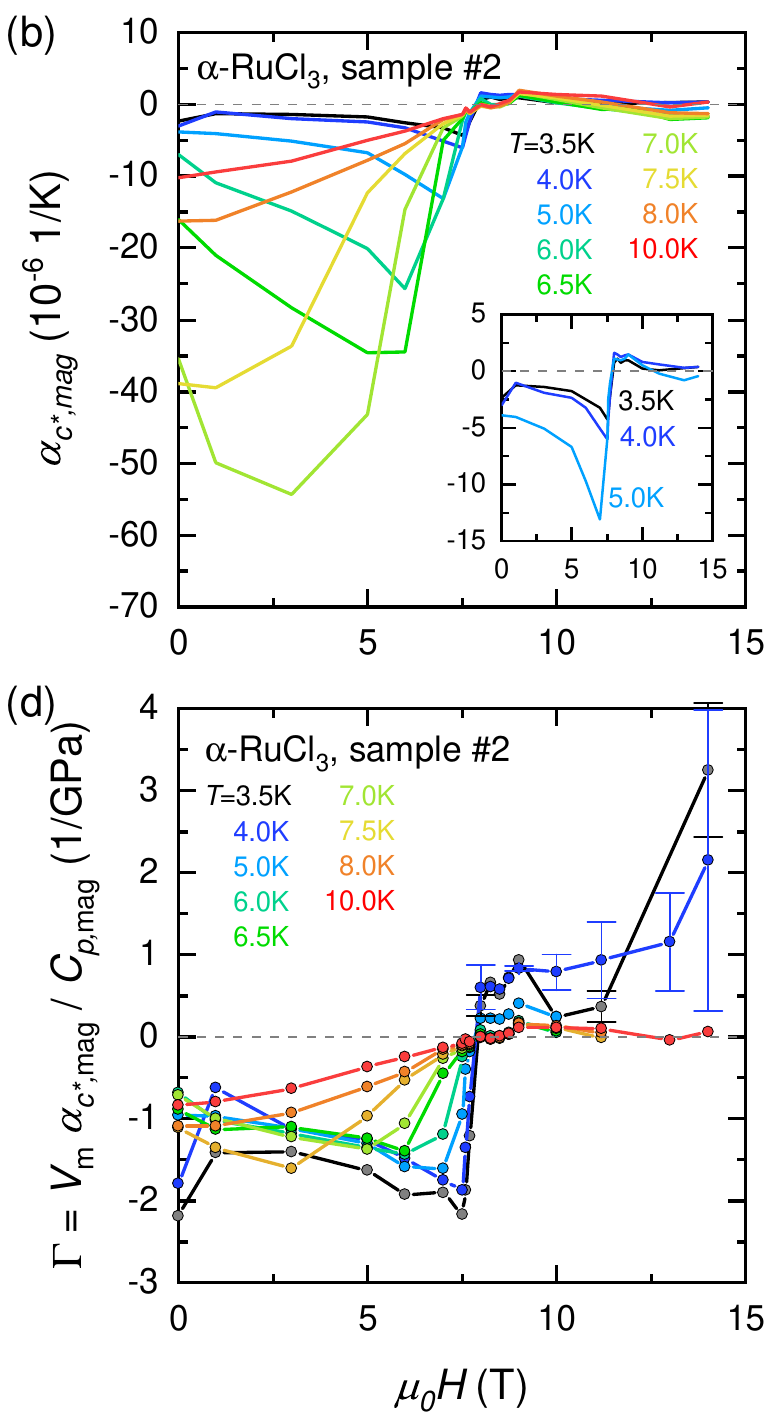}
\caption{\label{gamma_sample1-2}
Field dependence of (a,b) the linear TE coefficient $\alpha_{c^*,\rm mag}$ and (c,d) the Gr\"uneisen ratio $\Gamma$ for (a,c) sample \#1 and (b,d) sample \#2 in magnetic fields $\vec H \parallel$ Ru-Ru bonds for different temperatures between $3.5$\,K and $10$\,K.}
\end{figure*}

Fig.~\ref{gamma_sample1-2} shows the field dependencies of $\alpha_{c^*,\rm mag}$ and $\Gamma$ for samples \#1 and \#2, respectively, for different temperatures between $3.5$\,K and $10$\,K. Note that the measurement accuracy of the two observables is rather different: While the error bar of $\alpha_{c^*,\rm mag}(H)$ is mainly determined by the reproducibility of our setup and the uncertainty of the subtracted phononic background contribution to $\alpha_{c^*}$ (see below), the error of $\Gamma(H)=V_{\rm m} \alpha_{c^*,\rm mag}/C_{p,\rm mag}$ is influenced by more factors.
This can be seen in our data on both samples, where $\Gamma(H)$ displays a large scatter and large errors in the high-field regime above $11.2$\,T, where both quantities $\alpha_{c^*,\rm mag}$ and $C_{p,\rm mag}$ become small due to the spin excitation gap. Still, both data sets (sample \#1 and \#2) are in good agreement with each other, evidencing the sign change of the linear TE coefficient and of $\Gamma$ at the critical field $\mu_0H_{\mathrm c1}$. Also the anomalous asymmetric Gr\"uneisen ratio for fields around the critical field and the small absolute values for fields just above $H_{\mathrm c1}$ are fully reproduced on both samples. We note that Fig.~\ref{gamma}(b) of the main paper displays $\Gamma(H)$ at $4$\,K for both samples and at $10$\,K for sample \#1.

\section{Phonon background subtraction}\label{app:phonon}

\begin{figure*}[p]
\includegraphics[scale=1.15]{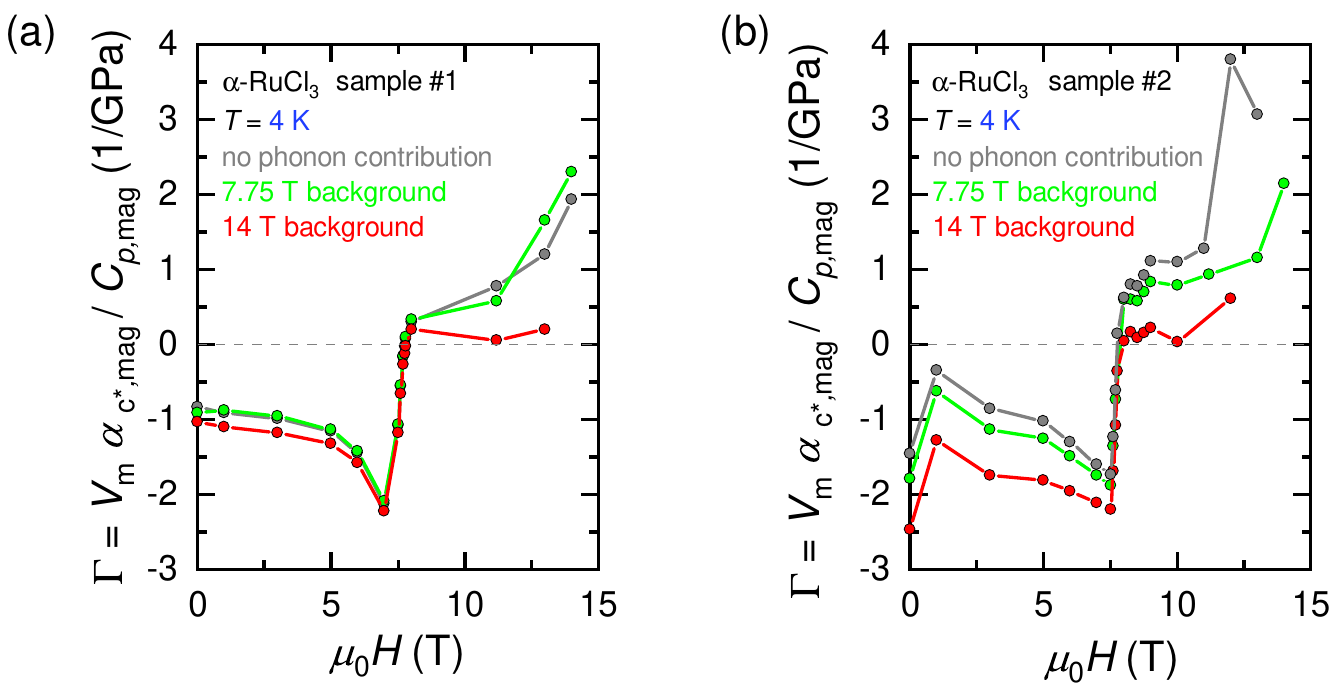}
\caption{\label{bkg_gamma}
The Gr\"uneisen ratio $\Gamma(H)$ of {\rucl} at $T=4$\,K as a function of the magnetic field $\vec H\parallel$ Ru-Ru bonds using different approximations for the phononic contribution to the TE for (a) sample \#1 and (b) sample \#2; for details see text.}
\end{figure*}

To check the sensitivity of the key features found in the Gr\"uneisen ratio with respect to the choice of the phonon background model for the TE coefficient, we tried two alternative approaches in addition to the scheme discussed in Sec.~\ref{subsec:exp-grueneisen}: (A) We modeled the phononic contribution to the TE using the $14$\,T TE data in the gapped high-field state. Although the spin excitation gap is still not large enough to fully gap out magnetic excitations above $\sim 6$\,K, the key features of the resulting Gr\"uneisen ratio at low temperatures are robust. (B) Since the phonon contribution to the TE is typically much smaller than that to the specific heat, we neglected this contribution in a first approximation for $T <$ 10~K. A comparison of $\Gamma (H)$ at $4$\,K for the different approaches is shown in Fig.~\ref{bkg_gamma}, emphasizing the robustness of the anomalous behavior of $\Gamma (H)$ for $H > H_{\mathrm c1}$ at low $T$.

\clearpage

\bibliography{aRuCl3_Paper_TEMS}

\end{document}